\documentclass[fleqn,12pt,twoside]{article}
\usepackage[headings]{espcrc1}

\readRCS
$Id: espcrc1.tex,v 1.2 2004/02/24 11:22:11 spepping Exp $
\usepackage{graphicx}
\usepackage[figuresright]{rotating}
\usepackage{psfrag}

\newcommand{\AmS}{{\protect\the\textfont2
  A\kern-.1667em\lower.5ex\hbox{M}\kern-.125emS}}

\title{Higher Twist Effects in Deep Inelastic Scattering on Nuclei}

\author{Hans J. Pirner\address[UHD]{Institut f\"ur Theoretische Physik 
                                    der Universit\"at Heidelberg,\\ 
                                    D-69120 Heidelberg, Germany}%
                ${}^,$\address{Max-Planck-Institut f\"ur Kernphysik,\\
                               D-69117 Heidelberg, Germany}
        and
        Daniel Gr\"unewald\addressmark[UHD]
}
       
\runtitle{Higher Twist Effects in Deep Inelastic Scattering on Nuclei}
\runauthor{Hans J. Pirner and Daniel Gr\"unewald}

\begin{document}

\maketitle

\begin{abstract}
Particle production in deep inelastic scattering on nuclei is reduced
due to absorption of the produced particles in the nucleus. 
The photon ejects a quark from a bound nucleon which propagates through 
the nucleus forming  a prehadron before turning into a hadron. 
We calculate the higher twist effect in hadronization which dominates the
$z \rightarrow 1$ region of fragmentation.
\end{abstract}

\section{Introduction}\label{intro}
Physics with nuclei extends the realm of strong
interactions from the hadronic size scale of 1 fm to the size scale
of nuclei of 10 fm. When a high energy probe illuminates the nucleus,
a picture of the nucleus develops which is different from 
nuclear spectroscopy or low energy nuclear physics.
Because of the high energy probe, QCD as the theory of quarks and
gluons is needed. On the other side  it is necessary to understand
strong
interactions at large distances. Three main issues are associated with hadronization 
in nuclei:
\begin{itemize}
\item At which distances  does color neutralize in the vacuum?
\item How can one see effects of prehadron formation?
\item Does the prehadron have a smaller cross section due to
      color transparency ?
\end{itemize}

Renewed interest arises from recent studies of deep inelastic
electron scattering on nuclei, where the produced hadrons are analyzed.
The HERMES \cite{HERM01,HERM03,HERM04,Muccifora02} experiment has
brought 
new insight into the question, since
it has access to
lower energy transfers $\nu$, where the hadronization
occurs inside the nucleus, and to a larger range of fractional momenta $z$
which determine the formation length of the hadron. The  nucleus helps to track
the space time evolution of a parton, since the nucleons
play the role of very nearby detectors of the propagating object. 
Deep inelastic electron nucleus scattering has the 
additional advantage that the
electron gives a well defined energy $\nu$ to the struck quark 
propagating through cold nuclear matter. The understanding of this
process is crucial for the interpretation  of  
ultra-relativistic proton-nucleus and nucleus-nucleus collisions.
Due to factorization in deep inelastic scattering the semi-inclusive
cross section can be described by the product of a parton distribution function
(PDF)
with  a fragmentation function (FF) cf. fig.~1.  
In the parton model the probability  
that a  quark
with flavor $f$ and momentum fraction $x$  
is present in the target is multiplied with  the probability 
that it hadronizes 
into a definite hadron which carries a momentum fraction $z$ of 
the quark.
\newline
fig.~1 shows a schematic diagram of semi-inclusive
deep inelastic lepton scattering (SIDIS) on a target $T$, and the definitions
of the four momenta of the particles involved in the process.
In SIDIS besides the scattered lepton $l\,'$
the leading hadron $h$ formed from the struck quark is detected 
with energy $E_h=z\nu$ in the target rest frame.
The summation over flavors includes the product of 
the fragmentation functions and structure functions for each flavor.
The experimental data on nuclear effects in hadron production 
are usually presented in terms of multiplicity ratios  as 
functions of $z$,$\nu$ or recently $Q^2$:
\begin{equation} 
    R_M^h(z)  = 
        \frac{1}{N_A^\ell} \frac{dN_A^h}{dz}
        \bigg/ \frac{1}{N_D^\ell} \frac{dN_D^h}{dz}. 
 \label{multrationu}
\end{equation}
In the above definitions $N_A^\ell$ is the number of outgoing leptons
in DIS processes on 
a nuclear target of atomic number $A$, while $dN_A^h/dz$ is
the $z$-distribution of produced hadrons in the same
processes; the 
subscript $D$ refers to the same quantities when the target is a
deuteron.  
In absence of nuclear effects the ratio $R^h_M$ would be equal to
$1$. In this paper we review 
the current status of the absorption model 
\cite{Accardi:2002tv,Accardi:2005jd} in describing 
hadron production in deep inelastic scattering and add recent results on higher
twist effects to semi-inclusive particle production at large $z \rightarrow 1$. 

\begin{figure}[t]
\begin{center}
\includegraphics[width=7cm]{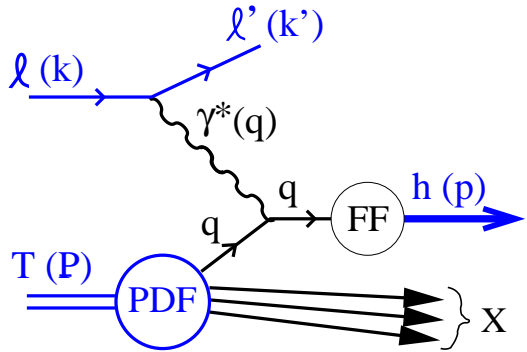}
\end{center}
\caption[]{Semi-inclusive hadron production in deep inelastic
scattering on a target T in the pQCD factorization approach. Parton 
distribution functions (PDF) and fragmentation
functions (FF) represent the non-perturbative
input.}
\label{fig:DIS}

\end{figure}

\section{Hadron Multiplicity Distribution}

The theoretical calculation takes into account several nuclear effects.
Parton distribution functions and fragmentation functions
both depend on the virtuality $Q^2$ of the DIS process.
This adjustment to  the scale $Q^2$ takes into account all 
radiated gluons before and after the photon quark interaction in 
the leading logarithm approximation. In nuclei gluon radiation may be affected
by the partial deconfinement of color which follows from overlapping
nucleons. 
Therefore in DGLAP evolution of 
nuclear structure and fragmentation functions the starting scale is smaller
and they evolve over a larger interval in momentum compared with 
the corresponding
functions in the nucleon at the same scale Q. 
Once color neutralization has taken place, gluon radiation stops as a
source of energy loss. A color neutral prehadron has been produced
which interacts with the color neutral other nucleons in the nucleus. 
In the work by A.~Accardi, V.~Muccifora, D.~Gr\"unewald  and
myself \cite{Accardi:2002tv,Accardi:2005jd}
similarly to the gluon bremsstrahlung model \cite{Kopeliovich:2004kq}
the time scale for prehadron formation  is short  for large $z$ and small $z$. 
It increases with z because of Lorentz time dilatation and
it is small for large $z$, since the hadron has to be formed
instantaneously, otherwise the energy loss downstream is too large.
\begin{figure}[htb]
\begin{center}
\vspace*{-5.0cm}
\includegraphics[width=13cm]{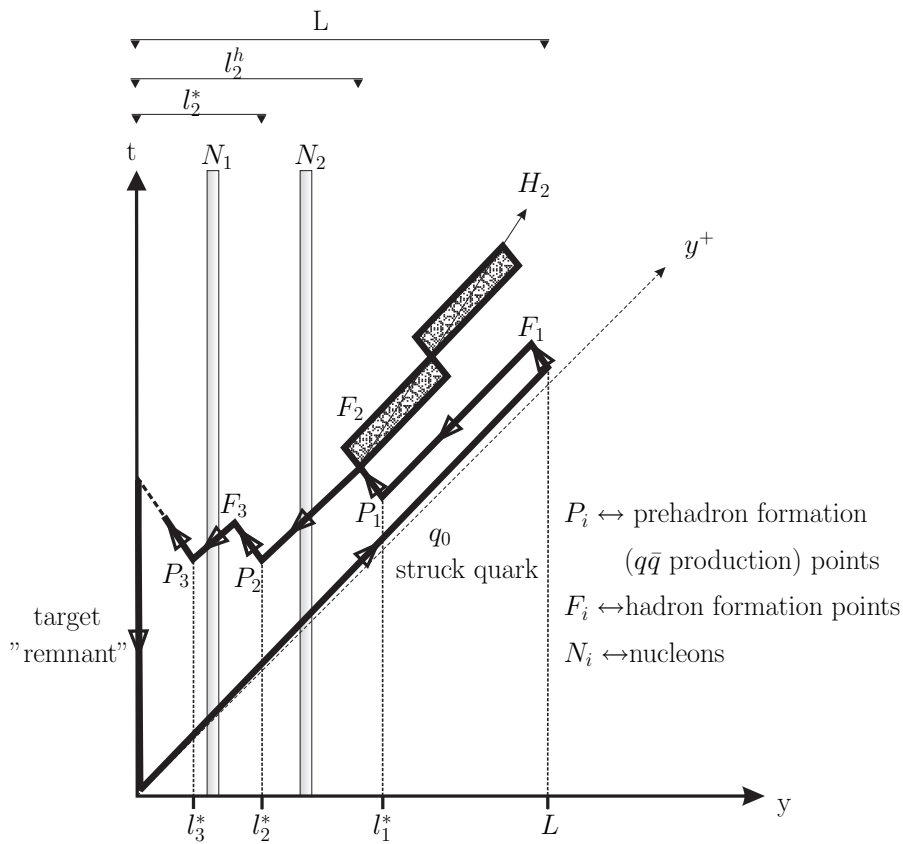}
\vspace*{-3.5cm}
\end{center}
\caption[]{Schematic space time picture of hadronization in the LUND model.}
\label{fig:lundmodel}
\end{figure}
In fig.~2 we show a schematic view of the hadronization process in the 
rest system of the nucleus. Our theoretical model follows closely
the Lund model for the fragmentation
process \cite{LundBook}. The nucleons $N_1,N_2,...$ are taken at 
rest in the nucleus. They are the target particles for the developing
fragmentation
cascade of the struck quark. We differentiate between the prehadron
formation points $P_i$ and the hadron formation points $F_i$.
The space-time development of the fragmentation process 
begins when the quark $q$ is
ejected from a nucleon. 
The quark propagates and the  colour string between the quark and the remnant
breaks into  smaller pieces. 
\newline
Hadrons are ordered according to their rank
$i$. Note that the first-rank hadron is always
created at the end after the original quark has traversed a
distance $L$
\begin{equation}
    L=\frac{\nu}{\kappa}\ ,
\end{equation}
where $\kappa=1$ GeV/fm is the string tension. 
This length $L$ can be very large
cf. fig.~2,
but this does not mean that long strings exist in the nucleus,
because the string breaks and prehadrons are formed.
There are two relevant lengths for the
fragmentation process,
the average position $\langle l_* \rangle$ at which the prehadron
is formed  and
the average distance $\langle l_h \rangle \leq L$ at which the hadron
is formed, where the averaging is over all hadron ranks.
In the Lund model the formation length of the prehadron and hadron are related
\cite{LundBook}:
\begin{equation}
    \langle l_h \rangle = \langle l_* \rangle + z L \ .
  \label{l*vslh}
\end{equation}
They both increase linearly with the
virtual photon energy $\nu$. However, as functions of $z$ they 
behave rather differently, especially at $z \rightarrow 1$, 
where $\langle l_* \rangle \rightarrow 0$ 
and $\langle l_h \rangle \rightarrow L$. 
\newline
Once the prehadron is produced, it makes inelastic collisions
which degrade its z fraction. Since the fragmentation function is very
steep, we consider the  prehadron lost after an inelastic collision.
We introduce
a survival probability  ${\cal S}_A(z,\nu)$ which
is calculated in linear transport equations which
include formation and absorption of the quark and prehadron.
These equations improve previous work 
by Bialas and Chmaj \cite{BC83} by treating hadronization
and absorption on equal footing.  
The survival probability 
represents the probability that  the prehadron 
and the hadron survives the traversal of the nucleus.
\newline
We calculate  the multiplicity distribution of a hadron h with
momentum fraction $z$ in a nucleus $A$ with the following expression
\begin{eqnarray}
  \frac{1}{N_A^{DIS}}\frac{dN_A^h(z)}{dz}& = &  
       \frac{1}{\sigma^{lA}} \hspace{-0.2cm}
       \int_{\mbox{\footnotesize exp. cuts}}
       \hspace{-0.6cm}
       dx\,d\nu \nonumber\\
   &&  \times \sum_f e_f^2 q_f^A(x,\xi_A Q^2) 
      \frac{d\sigma^{lq}}{dx d\nu} S_{f,h}^A(z,\nu) D_f^h(z,\xi_A Q^2).
\label{eq:SIDIS_def}
\end{eqnarray}
In the formula above $e_f$ is the electric charge of a quark of flavor $f$,
$d\sigma^{\gamma^* q}/dx\,d\nu$ is the differential cross section for
a $\gamma^* q$ scattering computed in pQCD at leading order.
The nuclear structure function $q_f(x,\xi_A(Q) Q^2)$ and the nuclear 
fragmentation function $D^h_f(z,\xi_A(Q) Q^2)$ are used. 
In ref. \cite{Accardi:2005jd} we fitted  a prehadron cross section equal to the
$2/3$ of the hadron cross section which is realistic since the prehadron coming
from gluon radiation of the struck quark is smaller than the
final hadron.

\section{Prehadron absorption}
The dependence of hadron attenuation on the mass number $A$ of 
the target nucleus is commonly believed  
to clearly distinguish the
absorption and energy loss mechanisms for hadron attenuation. Indeed,
the common expectation is that attenuation in absorption models is
proportional to the hadron in-medium path length $d$, leading to 
$R_M\propto A^{1/3}$. On the other hand, the average energy loss for
a parton traversing a QCD medium is proportional to $d^2$, which leads
to 
$R_M\propto A^{2/3}$. 

\begin{figure}[ht]
\begin{center}
\includegraphics[height=1.0\linewidth]{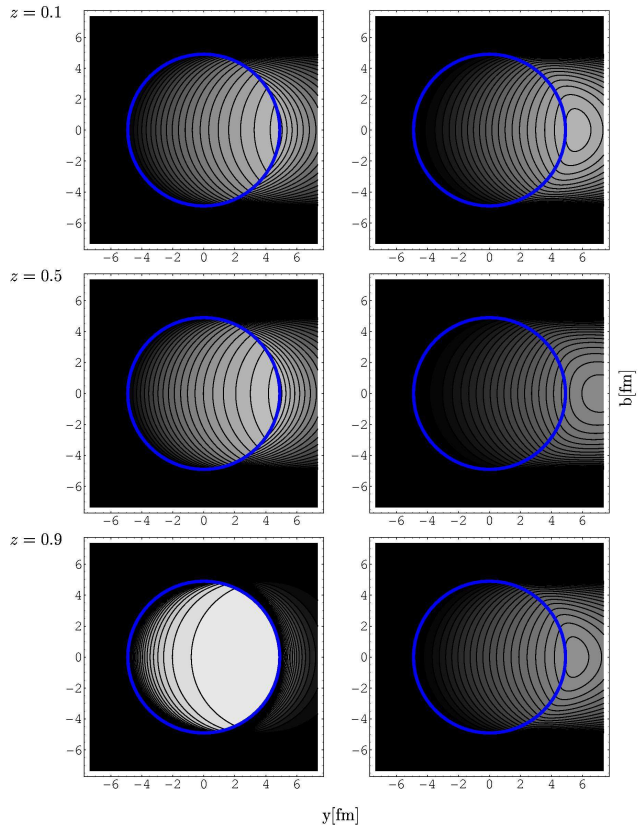}
\vspace*{-4.0cm}
\end{center}
\caption[]{Distribution of prehadrons and hadrons with
  $z=0.1,z=0.5,z=0.9$ in Kr induced by an
  interaction of a photon with a corresponding average energy $\langle \nu   
  \rangle(z)$ given by the HERMES \cite{HERM01,HERM03,HERM04,Muccifora02} experiment.}
\label{fig:FormPointDistrFinal}
\end{figure}
In fig.~3, we show a picture where the different formation areas of
a premeson (left part of the figure) and a meson (right part of
the figure) are shown relative to a target  Kr nucleus which has  a
radius of about 5 fermi. From top to bottom the figures represent different
$z$-values, 
$z=0.1,\,z=0.5$ and $z=0.9$
are shown. The energy of the initial quark for the different values of z is given by the respective average of the virtual photon energy $\langle \nu \rangle (z)$ given by the HERMES \cite{HERM01,HERM03,HERM04,Muccifora02} experiment.
One sees that the hadron dominantly appears outside of the nucleus, whereas
the prehadron forms inside the nucleus, especially for
large and small $z$-values.
Therefore we take the
model described in ref. \cite{Accardi:2003gp} in 
the one step (= prehadron)
approximation and neglect final hadron production in the
calculation of the survival probability.
To further simplify, we consider the case of a hard-sphere nucleus of
mass number $A$ and radius $R =  r_0 A^{1/3}$, with $r_0=1.12$ fm. 
Neglecting
absorption in deuterium and the small rescaling correction we find
that the hadron multiplicity ratio $R_M$ equals the hadron survival
probability $S_A$.
These simplifications make an analytical computation of the attenuation
 possible (cf. ref. \cite{Accardi:2005jd}) and an expansion of the result in terms of
the mass number yields 
 \begin{equation}
  1-R_M = c_1 A^{2/3}+c_2 A^{4/3} + \mathcal{O}[A^2]  \ .
 \label{form:survBC}
\end{equation}
The numerical values of $c_1$and $c_2$ depend on $z$ and $\nu$,
in general $c_2<<c_1$. 
A fit to the theoretical model without any simplifications and 
to the experimentally measured multiplicity ratios finds $1-R_m=c A^{\alpha}$
with a strong correlation between $\alpha$ and $c$ 
favoring $\alpha=2/3$ in agreement with the above considerations on the absorption model.

\section{Higher Twist Effect near $z \rightarrow 1$}
\begin{figure}
 	\begin{minipage}[t]{0.49\linewidth}
	  \psfrag{x1}{\scriptsize $z$}
	  \psfrag{x2}{\hspace{-0.05\linewidth} \scriptsize $R_M^{\pi^+}(z)$}
		\fbox{\includegraphics[width=0.97\linewidth]
		    {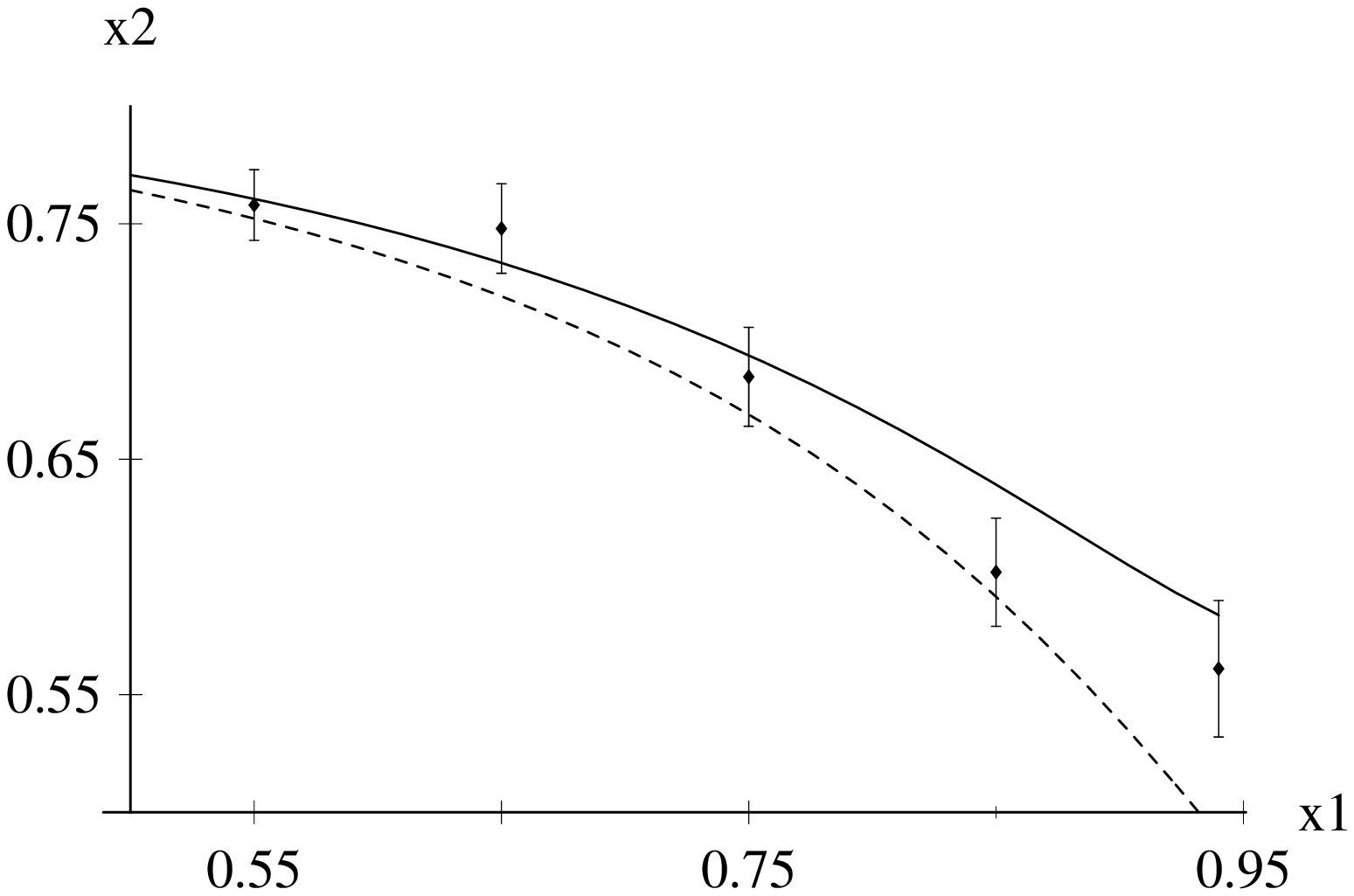}}    
	\end{minipage}
	\begin{minipage}[t]{0.49\linewidth}
		\psfrag{x1}{\scriptsize $z$}
	  \psfrag{x2}{\hspace{-0.05\linewidth} \scriptsize $R_M^{p}(z)$}
		\fbox{\includegraphics[width=0.97\linewidth]
		    {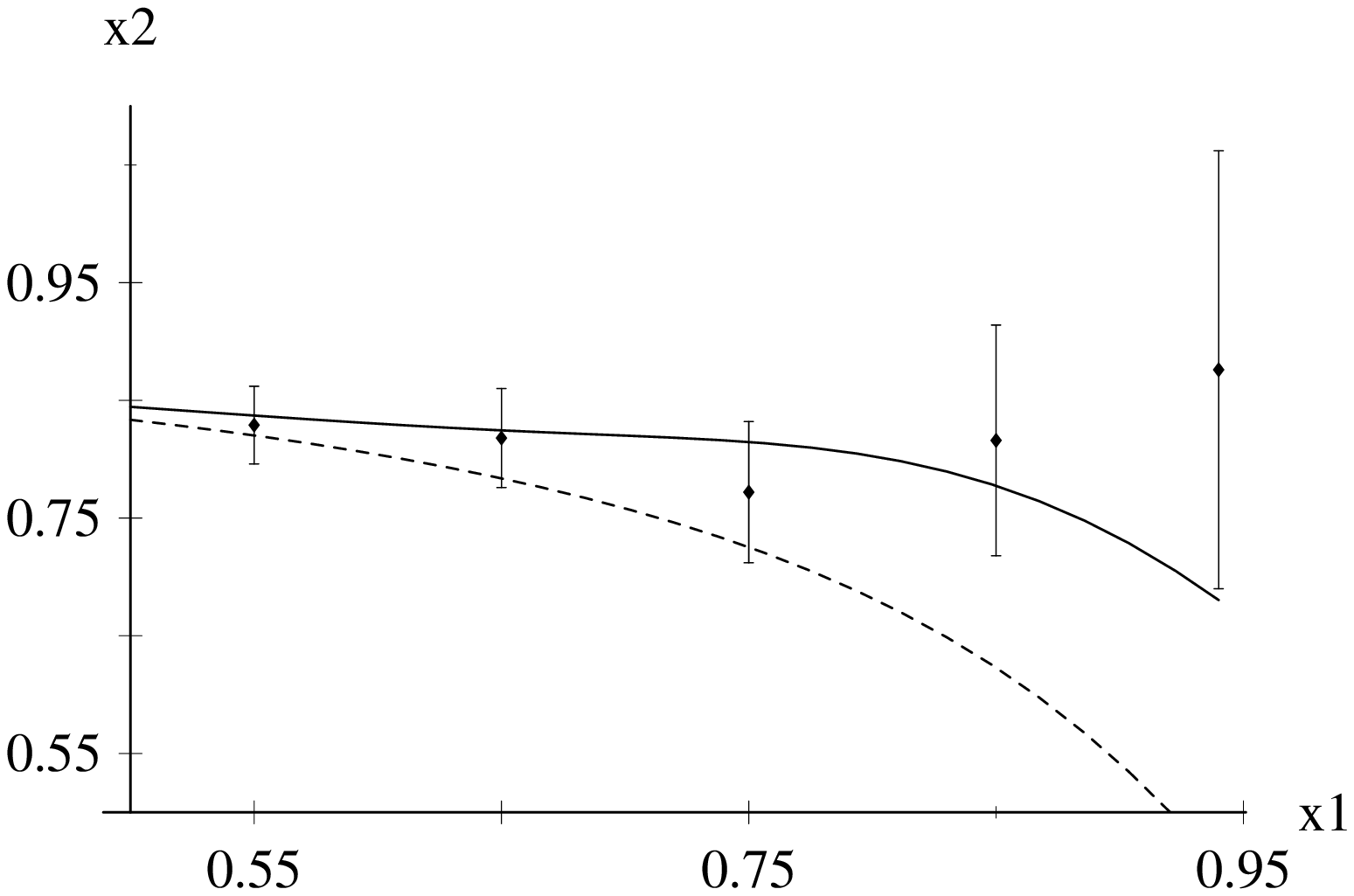}}
	\end{minipage}		
	\caption{Multiplicity ratios $R_M(Kr)$ for pions (left panel) and protons (right panel)
	         as a function of $z$. Dashed lines show the leading twist fragmentation process
	         of ref. \cite{Accardi:2005jd} computed in the one step approximation. The full 
	         lines show the modifications due to the higher twist process. The points show the 
	         experimental HERMES data \cite{HERM03}.
	}
	\label{fig:MPRatio_HT}
\end{figure}

In the late eighties E.L.  Berger  \cite{Berger:1979kz}
has published several papers on higher twist effects in semi-inclusive particle production.
The fragmentation function $D_{h/q}(z,Q²)$ contains a leading twist term  $D_{h/q}^{LT}(z,Q²)$ 
which evolves logarithmically with $Q^2$ due to perturbative gluon radiation and a higher
twist term $D_{h/q}^{HT}(z,Q²)$ responsible for direct hadron production
\begin{eqnarray}
D_{h/q}(z,Q^2) &=& D_{h/q}^{LT}(z,Q²)+D_{h/q}^{HT}(z,Q²).       
\end{eqnarray}
The higher twist contribution to the fragmentation function for production of pions is modeled 
in ref. \cite{Berger:1979kz} by:
\begin{equation}
D_{\pi/q}(z,Q^2)= D_{\pi/q}^{LT}(z,Q²)+\frac{1}{3} F_{\pi}(Q^2).
\end{equation}
where  $F_{\pi}(Q^2)$ is the pion monopole form factor 
$F_{\pi}(Q^2) \propto m_{\rho}^2/(m_{\rho}^2+Q^2)$ determined by the $\rho$ meson
mass. The leading twist fragmentation function at low $Q_0^2=2\,GeV^2$ behaves as  
$ D_{\pi/q}^{LT}(z,Q_0²)\propto (1-z)$. This means that for $z>0.7$ the higher twist direct
pion production takes over. 
The radiated gluon and remaining quark must carry each about one half of the  momentum
of the struck quark. The resulting propagators suppress the higher twist
contribution. 
\newline
For nucleon production, we expect an even larger higher twist suppression factor:
\begin{eqnarray}
D_{N/q}(z,Q^2) &=& D_{N/q}^{LT}(z,Q²)+D_{N/q}^{HT}(z,Q²)\nonumber\\
               &=& D_{N/q}^{LT}(z,Q²)+\frac{1}{3} F_{N}(Q^2).
\end{eqnarray}
The resulting nucleon form factor has a dipole behaviour. A comparison of our
previous calculations \cite{Accardi:2005jd} with the experimental data indicates 
that for large z-values an additional mechanism is at work. Therefore we have estimated that the attenuation of the directly produced higher twist hadron must be less ,
due to color transparency. We fitted a reduced cross section
$\sigma^{CT}=0.37 \sigma ^{\pi N}=10\,mb$ both for the direct pion and nucleon
production. In fig. \ref{fig:MPRatio_HT} we show the results for the multiplicity ratios
$R_M$. In both cases $R_M$ falls off more slowly for $z \rightarrow 1$. 
For the nucleon it more or less flattens out. 
\newline
This purely phenomenological study puts numerical estimates by Kopeliovich and 
Nemcik into a new light. Indeed for $z >0.7$ direct hadron formation is important
for small $Q^2$. Note the average $Q^2=2.5\,GeV^2$ in the Hermes data. 
The formalism to be developed may allow to calculate the color transparency effect
both for the leading twist prehadron and the higher twist produced direct hadron.
Deliberately we left out to parametrize the $Q^2$ dependence of the 
color transparency reduced cross section $\sigma^{CT}$, since the data on $R_M$ do not
need such a dependence. On the other side the semi-inclusive cross sections themselves
should show a strongly $Q^2$ dependent higher twist effect in the region $z>0.7$
and $Q^2<4\,GeV^2$. Higher twist fragmentation is additionally enhanced in the 
nucleus because of its smaller absorption cross section and thereby becomes
better accessible experimentally. 

\section{Conclusions}
In summary, the absorption model 
\cite{Accardi:2002tv,Accardi:2005jd} 
can describe  the HERMES data on the nuclear
modification of hadron production in DIS 
processes. A reasonable choice of parameters with the vacuum
string tension and a prehadronic cross section equal to $2/3$ of the
hadronic cross section explains the data. The $A$-dependence in the
absorption model is simple. 
The attenuation $1-R_M$ is proportional to $A^{2/3}$ since mostly prehadrons from the away surface of the nucleus contribute
to the cross section. 
\newline
Finally high energy electron-nucleus collisions 
allow to observe the higher twist effect in the fragmentation
function at $z \rightarrow 1$. The directly produced higher twist
hadrons have a  smaller prehadron cross section
$\sigma^{CT}=10\,mb$ due to color transparency.
The nucleus enhances this higher twist effect.

\section*{Acknowledgments}

We are very grateful to B.~Kopeliovich
for discussions.  
This work is partially funded by the EU program 
Hadron Physics (RII3-CT-2004-506078).


\end{document}